\documentclass[sigconf]{acmart}

\usepackage{amsmath}
\usepackage{todonotes}
\usepackage[T1]{fontenc}
\usepackage{tabulary}
\usepackage{hhline}

\usepackage{filecontents}

\newcommand{\figwidth}{1}
\newcommand{\figwidthsmall}{.95}

\newcolumntype{K}[1]{>{\centering\arraybackslash}p{#1}}
\newcommand{\lfm}{Last.fm~}
\newcommand{\lfmnsp}{Last.fm}
\newcommand\given[1][]{\:#1\vert\:}
\setcopyright{rightsretained}
\acmConference[MLG'17]{13th International Workshop on Mining and Learning With Graphs}{August 2017}{Halifax, Nova Scotia, Canada} 
\acmYear{2017}
\copyrightyear{2017}

\acmPrice{}
\acmDOI{}

\acmISBN{}
\makeatletter
\def\@copyrightspace{\relax}
\makeatother

\begin{document}

\title{Evaluating Social Networks Using Task-Focused Network Inference}

\author{Ivan Brugere}
\affiliation{%
  \institution{University of Illinois at Chicago}
  \city{Chicago} 
  \state{IL} 
}
\email{ibruge2@uic.edu}

\author{Chris Kanich}
\affiliation{%
  \institution{University of Illinois at Chicago}
  \city{Chicago} 
  \state{IL} 
}
\email{ckanich@uic.edu}

\author{Tanya Y. Berger-Wolf}
\affiliation{%
  \institution{University of Illinois at Chicago}
  \city{Chicago} 
  \country{IL}
 }
\email{tanyabw@uic.edu}

\begin{abstract}
Networks are representations of complex underlying social processes. However, the same given network may be more suitable to model one behavior of individuals than another. In many cases, aggregate  population models may be more effective than modeling on the network. We present a general framework for evaluating the suitability of given networks for a set of predictive tasks of interest, compared against alternative, networks inferred from data. We present several interpretable network models and measures for our comparison. We apply this general framework to the case study on collective classification of music preferences in a newly available dataset of the \lfm social network.
\end{abstract}

\maketitle

\section{Introduction}

Networks are used as representations of complex underlying social processes between individuals. In the context of social networks, researchers typically assume that the observed, boolean network (friend/not friend) is a suitable model to explain future behavior.  

Networks constructed from a social process (e.g. `friendship') are an accumulation of these expressed relationships, but to what extent do these relationships model the observed behavior on the network at any given time? Can we measure which edges are actually informative for explaining a set of behaviors or performing predictive tasks?

Our work evaluates the extent which a given network is the appropriate model of observed behavior in the user population. We robustly evaluate the extent of association between many observed behaviors of individuals and a fixed, observed network structure. Existing work infers parameters associated with correlations between network structure and label/attribute distributions. But this does not ensure the estimated model is actually \textit{useful} for the question of interest. Our approach is \emph{task-focused}, viewing the network as a representation for modeling a set of predictive tasks (e.g. node-label classification).  

We present alternate, interpretable models using networks inferred from data. Underlying node-attribute data in these networks corresponds to individual user activity such as logs of purchasing, product reviews, or media viewing/listening history. Networks from these data represent inferred relationships of shared or similar behaviors. These alternate models enable us to quantify--for example--to what extent `friends' predict each-other's music preferences vs. alternative global and local models from data. 

\section{Related Work}
Our work is related to statistical relational learning on \textit{attributed} and \textit{labeled} graphs. This work exploits statistical relationships between attributes, labels and network structure for several tasks: predicting new edges in a graph using attribute similarity \cite{Liben-Nowell2007, Hasan2011}, inferring missing attributes or labels given local attribute distributions in the network \cite{sen:aimag08}, or some combination of these \cite{namata:tkdd15,Gong:2014:JLP:2611448.2594455}.

Network \textit{structure} inference \cite{2016arXiv161000782B, Kolaczyk2014} uses statistical measures between attribute and label distributions of nodes to produce a network topology where none is explicitly given. These approaches are broadly applied in many domains, including computational biology \cite{ZhangHorvath2005}, neuroscience \cite{Sporns2014}, epidemiology \cite{Welch2011a} and recommender systems \cite{McAuley:2015:INS:2783258.2783381}. These approaches are typically in two categories: (1) model parameter inference under some assumptions, and (2) heuristics, expert tuning, or cross validation for particular tasks. 

Model-based approaches learn correlations between network structure and attributes or labels. These methods use parameter estimation and model selection to represent the network structure from data. These methods include the Attributed Graph Model (AGM) \cite{Pfeiffer:2014:AGM:2566486.2567993}, Multiple Attribute Graph (MAG) model \cite{Kim2012}, and Exponential Random Graph Model (ERGM) \cite{Robins2007}. 

In information networks, model inference is done on information arrival time of content over an unobserved network, which accounts for the rate or likelihood of transmission between entities \cite{Gomez-Rodriguez:2012:IND:2086737.2086741, Myers2010}. Challenges in this area include evaluating whether the inferred network is suitable for subsequent tasks such as classification, and how the model can yield \textit{alternative} representations for these tasks.

Task-focused approaches are a broad collection of methods which make few modeling assumptions. These models make global, local, or conditional thresholding choices on a relational or similarity space, e.g. given by a similarity measure over node pairs or simple interaction counts \cite{DeChoudhury:2010:IRS:1772690.1772722}. These approaches use predictive evaluation criteria such as cross validation against a particular \textit{task} of interest.  These methods often have a high sensitivity to threshold/parameter choice, and added complexity of interactions between network representations and task methods.

\section{Contributions}
We apply one instance of our general network structure inference framework\footnote{To appear.} for empirical analysis of the \lfm dataset. This framework uses local and global task methods to evaluate networks inferred from data in a general, modular, robust way across many observed behaviors. Our work makes the following contributions:
\begin{enumerate}
	\item We briefly introduce our existing general, attributed network testing framework to evaluate observed or inferred networks against interpretable network models, for arbitrary sets of tasks.
	\item We demonstrate evaluating an observed network against alternative models for collective label inference using our framework.
	\item We provide a detailed case study of a novel dataset collected from the \lfm social network. This dataset includes $4.6M$ users and $\approx 77B$ time-stamped song plays. We make this dataset available for further research.
\end{enumerate}
\section{Network Inference and Evaluation Framework}
\label{methods}

Our framework evaluates an observed network as one of several possible models for performing a task of interest (e.g. label prediction, link prediction). This allows us to evaluate how well an observed network explains the observed behavior of entities (e.g. users, individuals) against competing models.

%Problem \ref{p:networkinference} gives a concise specification of our model selection framework, including inputs and outputs. Given individual node-attribute vectors $\vec{a}_i \in A$, where `$i$' corresponds to node $v_i$ in node-set $V$, and a collection of node-labelsets $L \in L^*$, where $l_i \in L$ the label of node $v_i$ in a single labelset $L$, our general task-focused network inference framework evaluates a set of possible network models $M =\{\mathcal{M}_1...\mathcal{M}_n\}$  where each $\mathcal{M}_j: \mathcal{M}_j(A,L) \rightarrow E_j'$ produces a network edge-set $E_j'$.\footnote{Notation: capital letters denote sets, lowercase letters denote instances and indices. Primes indicate \emph{predicted} and inferred entities. Greek letters denote method parameters. Script characters (e.g. $\mathcal{C}()$) denotes functions, sometimes with specified parameters.} 

Our framework accepts an attributed, labeled network: $G = (V, E, A, L)$, with node $v_{i}$ in node-set $V$, edge $e_{ij}$ in edge-set $E$, an attribute vector $\vec{a}_i$ for node $v_i$ in node attribute-set $A$, and a collection of node label-sets $L\in L^*$ where $l_i \in L$ the label of node $v_i$ in a single labelset $L$.\footnote{Notation: capital letters denote sets. Lowercase letters denote instances and indices. Primes indicate predicted and inferred entities. Arrows denote vectors (e.g. $\vec{a_i}$. Greek letters denote method parameters. Script characters (e.g. $\mathcal{C}()$) denote functions, with specified parameters as needed for clarity.}

Nodes represent individual entities (e.g. users). Node attributes represent some high-dimensional activity history of the individual (e.g. product purchases, music listening/content viewing) on which we learn and evaluate `alternative' network definitions. A labelset represents a low-cardinality (e.g. boolean) attribute of predictive interest, which we use to learn joint relationships between attributes, edges, and labels.\footnote{Labels are node attributes, made distinct as $L$ only for notational convenience.} Such labelsets may include whether a user purchases a product or listens to a particular genre of music. Finally, edge-sets represent (1) observed, explicit edges in the social network, (2) inferred edges from attribute similarities or (3) the output of some parametric network model on attributes.

Our task-focused network inference framework evaluates an observed $G$ against a set of alternative network models $M =\{\mathcal{M}_1...\mathcal{M}_n\}$  where each $\mathcal{M}_j: \mathcal{M}_j(A,L) \rightarrow E_j'$ produces an edge-set. These networks (and the observed edge-set $E$) are evaluated on a set of inference task methods $C = \{\mathcal{C}_1...\mathcal{C}_m\}$ where each $\mathcal{C}_j: \mathcal{C}_j(E,A,L) \rightarrow P_j'$ where $P_j'$ is a collection of predicted edges, attributes, or labels depending on the task of interest. Finally, we evaluate predicted $P'$ under loss $\mathcal{L}(P, P')$, where $P$ is the validation or evaluation data.\footnote{For clarity, we refer to the network representation as a \textit{model}, and the task as a \textit{method}.}

An intuitive class of network models constructs edges by node-attribute similarity, for example the $k$-nearest neighbor network. This model associates a user with the top-$k$ most similar users according to purchase or review history, music listening history etc. according to the application. A specific task of interest is label prediction. In different applications this corresponds to predicting whether a user listens to a certain `genre' of music or buys a certain brand or type of product, inferred from their friends' behavior.  

This general framework requires specifying (1) network model-set $M$, (2) task method-set $C$, and (3) loss $\mathcal{L}()$ appropriate to the application. We instantiate a set of fundamental models and methods, applicable across numerous domains.

\subsection{Network models}
\label{subsec:nets}

Our framework accepts any appropriate models $\mathcal{M}: \mathcal{M}(A,L) \rightarrow E'$ for inferring structure from attributes and labels. 

Given a similarity function $\mathcal{S}(\vec{a}_i, \vec{a}_j) \rightarrow s_{ij}$ and a target edge count $\lambda$ we define two alternative edge-sets according to:

\begin{itemize}
	\item $k$-nearest neighbor $\mathcal{M}_{KNN}(A, \mathcal{S}(), \lambda)$: for a fixed $v_i$, select the top $\lfloor\frac{\lambda}{|V|}\rfloor$ most similar $\mathcal{S}(\vec{a}_i, \{A \setminus \vec{a}_i\})$. In directed networks, this produces a network which is $k$-regular in out-degree, with $k=\lfloor\frac{\lambda}{|V|}\rfloor$.
	\item Threshold $\mathcal{M}_{TH}(A, \mathcal{S}(), \lambda)$: for \emph{any} pair $(i,j)$, select the top $\lambda$ most similar $s_{ij}$. 
\end{itemize}
Evaluating these network models relative to an observed $G$ measures the extent that the \emph{relative} or \emph{absolute} similarity produces better performance on the task $\mathcal{C}()$. This choice between local and global network definition are a fundamental challenge across multiple domains. The $\mathcal{M}_{KNN}$ model is equivalent to setting a threshold on the pairwise similarity distribution \emph{per node} to produce equal degree. The $\mathcal{M}_{TH}$ model yields \emph{equal priority} of edges, ordered by the full pairwise similarity distribution.

We use a simple `intersection' similarity measure, which is suitable for comparing item count data as in our application. This is the sum of intersecting values over every attribute dimension. Given two attribute vectors $(\vec{a}_i, \vec{a}_j)$ with non-negative elements: 

\begin{equation}
	\mathcal{S}_{INT}(\vec{a}_i, \vec{a}_j) = \sum_k \mathrm{min}(a_{ik}, a_{jk}) 
\end{equation}

Although this is a naive similarity, we show it is interpretable and effective to quantify network comparisons. By design, the $\mathcal{M}_{TH}$ network model under $\mathcal{S}_{INT}$ is biased toward dense vectors with larger attribute magnitude, because $\mathcal{S}_{INT}$ uses absolute aggregation of counts. Higher similarity may correspond to more active nodes, which measure high intersection with other active nodes simply by magnitude. %In $\mathcal{M}_{KNN}$ low-activity nodes likely preferentially attach to high-activity nodes or specialized. 

In our application, this bias corresponds to creating associations between nodes with \emph{more robust} attribute distributions. In a real application, these robust nodes could be \textit{transferred} to evaluating the sparser attribute vectors of disconnected nodes under the network model. In the below application of \lfmnsp, $\mathcal{M}_{TH}$ under $\mathcal{S}_{INT}$ disconnects $60\%$ of nodes. We ignore any `coverage' measure (e.g. recall) to focus on task performance.

\subsection{Classification task methods and network evaluation}
\label{subsec:class}
Our framework evaluates the above inferred network models as well as observed networks against some set of tasks. We focus on collective node classification tasks. Given edge-set $E$, a neighborhood function $\mathcal{N}(E, i)$, and node-attribute set $A$, our node classification task infers unknown label $l_i$ from attribute \emph{test} vector $\vec{a}_i$ with a method trained on \emph{neighborhood} attributes and/or labels of $v_i$: $\mathcal{C}(A_{\mathcal{N}(E,i)}, L_{\mathcal{N}(E,i)}, \vec{a}_i) \rightarrow l_i'$. 

Our focus is primarily methodological and empirical, so we apply standard existing methods:
\begin{enumerate}
	\item Random Forest: $\mathcal{C}_{RF}(A_{\mathcal{N}(E,i)}, L_{\mathcal{N}(E,i)}, \vec{a}_i)$
	\item Linear Regression: $\mathcal{C}_{LR}(A_{\mathcal{N}(E,i)}, L_{\mathcal{N}(E,i)}, \vec{a}_i)$
	\item Naive Bayes: $\mathcal{C}_{NB}(A_{\mathcal{N}(E,i)}, L_{\mathcal{N}(E,i)}, \vec{a}_i)$
	\item Max-median Similarity: $\mathcal{C}_{CS}(A_{\mathcal{N}(E, i)}, L_{\mathcal{N}(E,i)},\mathcal{S}(), \vec{a}_i)$
	\item Neighborhood Most-frequent Label: $\mathcal{C}_{NL}(L_{\mathcal{N}(E,i)})$
	\item Global Attributes baseline: $\mathcal{C}_{GA}(A, L, \vec{a}_i)$
	\item Global Label Distribution sample baseline: $\mathcal{C}_{GL}(L)$
\end{enumerate}

The Random Forest, Linear Regression, and Naive Bayes methods all learn discriminating features on the attribute vectors $\vec{a}_j \in A_{\mathcal{N}(E,i)}$. The Max-median Similarity method builds distributions of similarity measures $\mathcal{S}(\vec{a}_i, \vec{a}_j)$, where $v_j \in \mathcal{N}(E,i)$ grouped by unique $l_j \in L_{\mathcal{N}(E, i)}$ values. The method reports the label with the maximal median of similarity. This measure only finds similarity across the entire attribute vector, rather than learning discriminative attribute dimensions.\footnote{Where convenient, we'll refer to network models or classification methods by their subscript, e.g. `KNN' rather than $\mathcal{M}_{KNN}$}

The Neighborhood Most-frequent Label uses the neighborhood function and label data, but does no learning on neighborhood attribute vectors. The Global Attribute method is a global \textit{baseline} which trains the same method as the corresponding neighborhood method, over all attribute vectors.\footnote{In practice, we evenly split attribute data $A$ and train two `global' methods. To test an $a_i$, we use the alternate method not trained using this data.} To compare ad-hoc local methods which do no explicit training (CS and NL), we return a global baseline estimate using the \emph{mean} performance of all other available GA methods. Finally, the Global Label Distribution samples labels according to global empirical frequency. In the binary case, this is a Bernoulli trial with probability $p=|L|$.

This set of classification methods allows us to measure task performance on each network model over varying attribute, label, and edge data availability. For example, comparing against the most competitive baseline, GA, evaluates whether the network is an appropriate model for the label classification task at all. If the global method performs better, then either we haven't used the appropriate network model to learn the local (neighborhood) heterogeneity in attribute-label correlations, or there are actually globally-discriminative attributes which are learnable across all nodes. In this case, the network isn't adding any information with respect to the attribute-label joint distribution.
\begin{table*}[ht]
	\centering
	\begin{tabular}{|c|c|c|K{1.2cm}|c|K{.9cm}|K{1.3cm}|K{.9cm}|c|c|}
		\hline
		Network & |V| & |E| & Median Degree & $\alpha$ & Median Age & Median Plays & Median Artists & Plays\footnote{estimated from aggregate user profile statistics collected prior to listening data collection} & Unique Artists\\\hline
		\lfm & 4,635,993 & 67,671,903 & 3 & 2.88 & 8.18Y & $\approx 3804$ & -- & $>77$B & -- \\
		\lfm $G_{200K}$ & 199,943 & 9,837,627 & 17 & 3.20& 7.75Y &32,628&461& 10,174,847,323 & 10,115,553 \\ 
		\lfm $G_{20K}$ & 19,990 & 678,761 & 35 & 3.42 & 7.25Y &41,454 &578& 1,243,483,909 & 2,866,276 \\
		\hline
	\end{tabular}
	\caption{A summary of the \lfm dataset: the full connected-component, and connected samples containing 200K and 20K users. Columns 2-5 report social network measures, Columns 6-8 report individual user statistics, Columns 9-10 report population statistics across users. Column 5 (`$\alpha$') refers to the Power-law exponent of the degree distribution, Column 6 (`Median age') measures the length of time since the user account was opened (without consideration of inactivity). Note that \lfm $G_{20K}$ and \lfm $G_{200K}$ contain very active users compared to the full dataset, with respect to the social network (`Median Degree') and listening activity (`Median Plays'). All datasets available at: \cite{data}.} 
	\label{tab:data}
\end{table*}

\section{Application Study: \lfm}
\label{application}
The \lfm  social network is a platform focused on music logging, recommendation, and discussion. Users can make `friends' in the network, which allows users to more easily track others' listening, and discuss on their profile pages. Users can also log song plays through various desktop and mobile players, which produces a time series log of song, artist and other meta-data (e.g. genre) per user.

We collected the largest connected component and all associated play logs of the \lfm social network as of March 2016, seeded from a very active account opened in 2006. This yielded $4.6M$ users with $67.6M$ edges, a median degree of $3$, and over $77B$ plays (see Table \ref{tab:data}). Users are ordered by breadth-first search from the seed, i.e. in non-decreasing geodesic distance. On this dataset, we sample the first $200K$, and $20K$ users to generate two datasets. The smaller subset yields a connected component with $678K$ edges, a median degree of $35$, and over $1B$ plays collectively by the users. Considering users of this dataset have $\approx 12x$ the friends as the median user and $\approx 11x$ the plays, these are especially active users.\footnote{To discover unseen users and/or another large connected component, we extracted users from the \href{https://www.last.fm/dashboard}{last.fm/dashboard} page, which reports users ``listening now.'' However, this yielded few positive results--very small connected components or disconnected nodes--which we omit from the dataset.} To demonstrate our framework, we focus on our smallest subset.

\subsection{Problem Introduction}

Do \lfm users ``pay attention'' to the social graph? Users may make few friends and rarely visit their friends' pages or interact through messages. These users may instead use the application primarily to individually log their music and receive recommendations from the \lfm recommendation engine. In this case, edges may represent a single user action years ago with no effective association to how music is socially shared. How can we evaluate whether the \lfm social network as expressed by users is a \emph{good model} for learning more detailed musical preferences than we could in an aggregate recommendation engine? We focus on ``genre listener identification'', an application of collective classification where we learn to classify rare-class users who listen to a genre from a hidden artist-genre association graph.

%In what instance might constraining the classification task to the network yield better results? Since there is more user data in aggregate, it seems a global classifier may always be preferable. For a given genre, certain subsets of that genre's artists may be more prevalent in local network clusters. This heterogeneity may be especially expressed in declared `friendship' edges , which are appropriately expressed . However, in aggregate these discriminative artists are smoothed out. Our framework However, when heterogeneous artist-genre mappings exist in different parts of the network are When heterogeneous  

%Measuring the ability of this given network for identifying rare-class, how musical `friends' We aim to learn a network which can We evaluate our network based on its performance at performing ``Genre listener identification''  This task classifies positive cases of listener of a genre. We focus for learning positive, rare-class labels from attributed network. In this application, each node's \emph{attributes} consist of a sparse vector, where each dimension stores the count of unique artist plays by that user.  

\subsection{Genre Labels}

We use the \lfm API to collect associations between $40$ user-generated genre tags, and artists with the top-$1000$ instances of each tag. There is significant noise in \lfm genre tags, especially for low-frequency tags. Therefore, top-tagged artists are hand-verified to confirm they are reasonable. The specific choice of genres is hand-picked for tag cleanliness, and intended to have varying global frequency and local clustering within the network. 

Our analysis focuses on artist plays, which is a more intuitive level of aggregation than song plays to compare user similarity. A user is a \emph{listener} of an artist if they've listened to the artist for a total of at least $5$ song plays. 

A user is a \textit{listener} of a genre if they are a listener of at least $5$ artists in the genre's top-$1000$ artists. A labelset $L$(`genre') $: l_i \in \{0,1\}$ is generated by the indicator function for whether user $v_i$ is a listener of `genre.' Users on the $G_{20K}$ network are at median a listener of 6 of 40 genres, and 15 of 40 genres at the $90$th percentile, with $1236$ users not a listener of any collected genre.

\begin{figure}
\centering
\includegraphics[width=\figwidth\columnwidth]{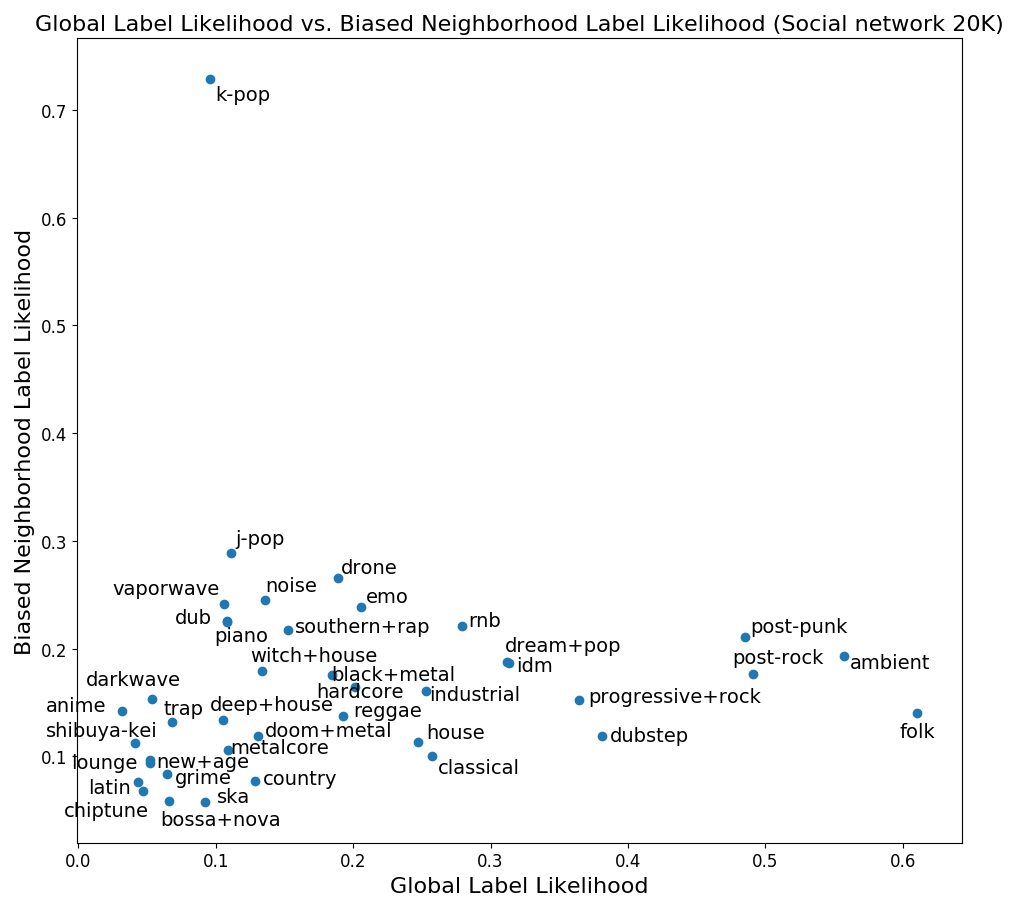}
\caption{The prevalence of genre listeners $|L|$, ($x$-axis) vs. the median of the network-label bias of co-listeners (Equation \ref{eq:bias}) on the $G_{20K}$ social network, ($y$-axis). Left to right measures obscure to common genres, down to up measures uniform to clustered genres with respect to network structure. For example, over half the users in the \lfm $G_{20K}$ dataset are listeners of `ambient' and `folk', while the neighbors of `k-pop' listeners are $\approx 70\%$ more likely to be `k-pop' listeners than expected by global listener prevalence ($\approx 10\%$).}
\label{fig:scatter_explicit}
\end{figure}

\subsection{Genre Label Empirical Analysis}

Figure \ref{fig:scatter_explicit} shows the global likelihood of the listener label per labelset, $|L|$ ($x$-axis) vs. the \textit{bias} of adjacent co-listeners ($y$-axis), i.e. over all listener nodes, the median of neighboring `listener' prevalence, minus the unbiased estimate:

\begin{equation}
\begin{aligned}
	\mathcal{B}(E, L): &~\mathrm{median}((|L_{\mathcal{N}(E,i)}| \given l_i=1) - |L|), \forall l_i \in L
\end{aligned}
\label{eq:bias}
\end{equation} 

This measures the local clustering of listeners in neighborhood $\mathcal{N}(E, i)$ per each listener node, summarized by the median, where $\mathcal{B}(E, L)=0$ indicates listeners are unbiasedly distributed on the network.   

Figure \ref{fig:scatter_explicit} shows an anomalous bias for `k-pop' listeners. This may be representative of user `friending' norms within this genre community, or an artifact due to our $G_{20K}$ network sample. Investigating further, this anomaly disappears in the $G_{200K}$ network. We write the \textit{change} in network-label bias from $E_{20K}$ to $E_{200K}$ as $\Delta\mathcal{B}(E_{200K},$`k-pop'$) = -0.52$. For notational convenience, we assume $\Delta\mathcal{B}$ from $E_{20K}$. In the larger sample, `k-pop' is adjusted to $\mathcal{B}(E_{200K},$`k-pop'$)= 0.21$, the $10$th most biased labelset of $40$ on $E_{200K}$. 

\begin{figure}
\centering
\includegraphics[width=\figwidth\columnwidth]{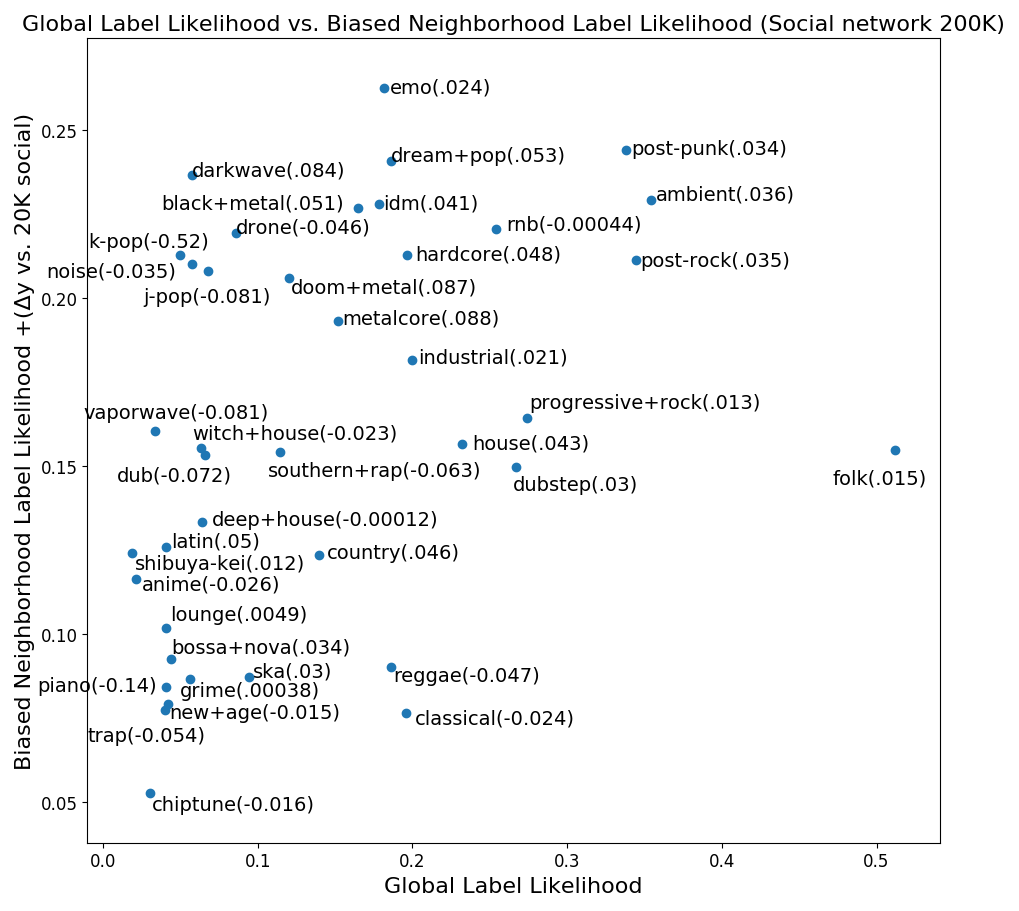}
\caption{The prevalence of genre listeners $|L|$, ($x$-axis) vs. the median of the network-label bias of co-listeners (Equation \ref{eq:bias}) on the $G_{200K}$ social network, ($y$-axis). Values in parentheses show the change in $y$-axis, i.e. $\Delta\mathcal{B}$ from $E_{20K}$ to $E_{200K}$. Note that most $\Delta\mathcal{B}$ values are small, except a large negative change in `k-pop.' This suggests these network-label bias statistics are robust in the $G_{20}$ sample.}
\label{fig:scatter200K}
\end{figure}

Figure \ref{fig:scatter200K} shows the global listener label prevalence vs. network bias on the $G_{200K}$ network, summarizing $\Delta\mathcal{B}$ in parentheses. Excluding `k-pop', the distribution of $\Delta\mathcal{B}$ over other labelsets (e.g. over all values in Figure \ref{fig:scatter200K} parentheses) is quite robust, summarized as $(\mu, \sigma)= (0.004, 0.050)$. $\mathcal{B}(E_{20K}, L)$ is therefore an unbiased estimate of $\mathcal{B}(E_{200K}, L)$ over $L^*$. This is evidence that the $G_{20K}$ sample is representative other active-user \lfm network samples with respect to label distributions, and subsequent results evaluated against these statistics are likely to generalize.

\begin{table}[ht]
	\centering
	\begin{tabular}{|K{2cm}|c|c|c|}
		\hline
		$\Delta\mathcal{B}: (\mu, \sigma)$& $E_{200K}$ & $E'_{KNN}$ & $E'_{TH}$\\\hline
		$E_{20K}$ & (0.004, 0.050) & (0.302, 0.124) & (0.183, 0.093) \\
		\hline
	\end{tabular}
	\caption{The $(\mu, \delta)$ of the distribution of $\Delta\mathcal{B}$ over labelsets, i.e. the values in parentheses in Figure \ref{fig:scatter200K}, from $E_{20K}$ (left), to edge-set (top). Note that $G_{20K}$ is an unbiased estimator for the network-label bias of $G_{200K}$ ($\mu=0.004$), and that $E'_{KNN}$ introduces $40\%$ more bias than $E'_{TH}$ with respect to the $E_{20K}$ social network edge-set.}
	\label{tab:labelsummary}
\end{table}

Table \ref{tab:labelsummary} summarizes $(\mu, \sigma)$ of $\Delta\mathcal{B}$ from $G_{20K}$ to each different network model. This quantifies the extent that our similarity measure $\mathcal{S}_{INT}$ `finds' co-listeners when constructing the alternative network. For $E'_{KNN}$ and $E'_{TH}$, we omit the analogues to Figure \ref{fig:scatter200K} for brevity, in favor of the Table \ref{tab:labelsummary} summary. 

To look at a few examples, several labelsets have a large gain in bias on the inferred networks:  $\Delta\mathcal{B}(E'_{KNN}, $`country'$)= 0.48$ , and $\Delta\mathcal{B}(E'_{KNN}, $`house'$)= 0.43$. This suggests that users are not building the \lfm social network on these shared interest, but that co-listeners of these genres \textit{can} be discovered in the population using the KNN network model. In contrast, some labelsets introduce little further bias: $\Delta\mathcal{B}(E'_{KNN}, $`vaporwave'$)= 0.01$, and $\Delta\mathcal{B}(E'_{KNN}, $`anime'$)= -0.036$. This suggests the social network is already representative of associating listeners by artist play similarity.

Furthermore, Table \ref{tab:labelsummary} shows that \textit{local} association via KNN introduces on average $\approx 40\%$ more bias than TH. This suggests that more active users are associated with other active users with a lower likelihood of co-listenership, relative to KNN under $\mathcal{S}_{INT}$. The KNN network model encourages associations among specialized communities with lower play counts. Quantifying this bias with respect to network models and different similarity measures allows further testing of the local and global properties of attributes, labels, and inferred network structures.

% \begin{figure}[h]
% \centering
% \includegraphics[width=.99\columnwidth]{figs/scatter_global}
% \caption{Measures ($x$-axis) the prevalence of listeners $P(l_i=1 \in L)$ vs. (y-axis) the median of relative prevalence of listeners on the observed network $\mathrm{median}(P(l_j=1 \in \mathcal{N}(i)|l_i=1)) - P(l_i=1 \in L)$. Left to right measures obscure to common genres, Down to up measures diffuse to clustered genres.}
% \label{fig:scatter_explicit}
% \end{figure}

%Figure \ref{fig:scatter_th} visualizes these statistics on $E'_{TH}$\footnote{$E'_{TH}$ omitted for brevity, summarized in Table \ref{tab:labelsummary}}, where each parentheses denotes ($\Delta\mathcal{B}$) from $G_{20}$ to $E'_{KNN}$ for `this' label-set. This measures how each network inference model affects the joint edge-label distribution, over many individual label-sets. Table \ref{tab:labelsummary} summarizes $(\mu, \sigma)$ over this distribution of $\Delta\mathcal{B}$. This shows that the $\mathcal{M}_{KNN}$ model introduces more bias. That is, users find more `listeners' similar to them when afforded the edges\todo{continue}

%\subsection{Inferred Network Local Clustering}

%As described in Section \ref{subsec:nets}, we define two alternative networks from pairwise comparison of attributes $(a_i, a_j) \in A$

\section{Evaluation}

We evaluate our observed, attributed, labeled $G_{20k}=(V,E,A,L^*)$ \lfm graph, as described in Section \ref{application} against $E'_{KNN}$ and $E'_{TH}$ network models described in Section \ref{subsec:nets}. 

\subsection{Genre Listener Classification}

We instantiate a collective classification problem for \textit{listener classification}. As an oracle, we provide local and global classification methods (see Section \ref{subsec:class}) the \textit{positive} genre-listener instances for testing, and their associated training attributes and labels under the neighborhood function for training. We focus on only positive instances because we want to learn the true rather than null association of genre `listener,' and positive labels are a rare class:  $\textrm{median}(|L|, L \in L^*) = 0.13$, therefore null instances would dominate the evaluation. Using multiple labelsets allows us to better evaluate the network model in the rare-class setting. %Rather than spending resources learning methods for non-listeners, we validate each network model over numerous rare-class label-sets.

For a given labelset $L$, each method produces $p'_i \in P'$ a set of node-label predictions (where $l_i=1$ is always the correct response under our oracle). We report the lift in precision relative to a baseline: $P'_b$:

\begin{equation}
    \mathrm{lift}(P', P'_b, L) = \frac{|P'| - |P'_b|}{|L|}
\end{equation}

\begin{figure}[ht]
\centering
\includegraphics[width=\figwidthsmall\columnwidth]{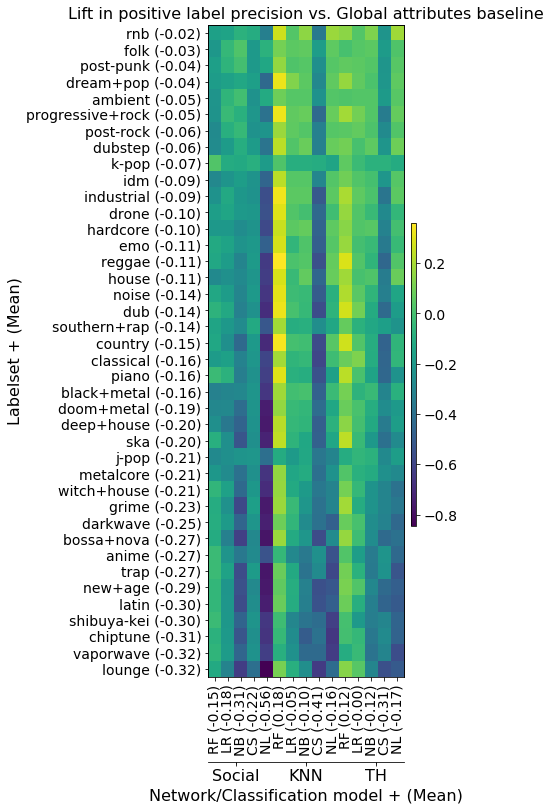}
\caption{For each model-classifier pair, the lift in precision against the Global Attribute Distribution baseline in classifying positive labels,  $l_i=1$ for $l_i \in L$, over each labelset. Row values in parentheses summarize \textit{mean} lift in precision for the labelset, and in columns summarize mean lift in precision for a model-classifier pair. This mean aggregation is also used for descending row-sort order.}
\label{fig:heatmap}
\end{figure}

Figure \ref{fig:heatmap} reports the lift in precision per labelset of model-classifier pairs (e.g. KNN-RF, TH-LR) vs. its Global Attribute (GA) baseline. Values in parentheses report the \textit{mean} lift over the row (for a labelset) or column (for a model-classifier pair). Figure \ref{fig:heatmap} shows there is not a labelset that outperforms the GA baseline for all model-classifier pairs. Visually we see that Random Forest (RF) performs best over each of the three network models. The $k$-Nearest Neighbor Random Forest (KNN-RF), and Threshold Random Forest (TH-RF)  outperforms GA on $35$ and $37$ of $40$ labelsets, respectively.

\begin{table}[ht]
	\centering
	\begin{tabular}{|c|c|c|c|c|c||c|}
		\hline
		$\mathcal{M}$ - baseline~\textbackslash~$\mathcal{C}$ &RF & LR & NB & CS & NL & \textbf{AVG} \\\hline
		Social-GL  & 0.06 & 0.27 & 0.33 & \textbf{0.43} & 0.09 & 0.24 \\
		KNN-GL & 0.39 & 0.41 & \textbf{0.55} & 0.24 & 0.49 & \textbf{0.42} \\
		TH-GL & 0.40 & 0.39 & \textbf{0.49} & 0.30 & 0.43 & 0.40\\
		Social-GA & \textbf{-0.15} & -0.18 & -0.31 & -0.22 & -0.56 & -0.28 \\
		KNN-GA & \textbf{0.18}& -0.05 & -0.10 & -0.41 & -0.16 & -0.11 \\
		TH-GA & \textbf{0.12} & -0.00 & -0.12 & -0.31 & -0.17 & \textbf{-0.10} \\		
		
		\hline \hline
		\textbf{AVG} & 0.14 & 0.01 & \textbf{0.16} &0.02 &0.14 & 0.09 \\
		\hline
	\end{tabular}
	\caption{The mean over labelsets of lift in precision on model-classifier pairs, against the Global Label (GL) and Global Attributes (GA) baselines. Network models: \lfm social network (20K), inferred KNN network (KNN), inferred Threshold network (TH). Classification methods: `RF' Random Forest, `LR' Linear Regression, `NB' Naive Bayes, `CS' Max-Median Cosine Similarity, `NL' Neighborhood Most-frequent Label.}
	\label{tab:results_full}
\end{table}

The Social Network Neighborhood Most-frequent Label (Social-NL) performs particularly poorly against both baselines. This means a user's neighborhood rarely yields a majority of co-listeners for a given label. This was also supported by Figure \ref{fig:scatter_explicit}, although now NL is further penalized by the performance of the GA baseline.

The mean values of model-classifier pairs (i.e. values in Figure \ref{fig:heatmap} column parentheses) are summarized in Table \ref{tab:results_full}, under both the GL and GA baselines. The GL baseline is very naive across all model-classifier pairs. However, GL highlights that Social Network Max-Median Cosine Similarity (Social-CS) is best under GA, and under GA it performs much closer to Social-RF and Social-LR than inferred models with CS against those same classifiers. This smaller performance gap suggests there is less discriminative learning on the `Social' neighborhood attributes, and whole-vector attribute similarity performs better than expected. 

Across all $5$ classification methods, `Social' has a positive lift in precision vs. GA for only $2$ labelsets total (`kpop', `folk') over $200$ label-classification pairs (i.e. all `Social' cells in Figure \ref{fig:heatmap}). This supports the hypothesis that the \lfm social network as declared by users isn't a good model for learning local music preferences. We will further quantify this in Section \ref{subsec:comparison}. 

There are several limitations of our results. This study evaluates the social network `as is', while several pre-processing steps and alternative evaluations could be suitable. For example, we do not prune abandoned accounts or impose other time-constraints. Weighted network models from attribute similarity could also be constructed, constrained to the social network.

However, all of these decisions evaluate a \textit{different} network object on a different property. Our intention is evaluating ``the network'' as given by the output of a social process of declared relationships, under minimal further assumptions. Often a researcher will be given ``the network'' under the assumption that its structure represents relevant relationships between individuals (often without abundant attribute data to verify). We focus on a general methodology for evaluating arbitrary networks in a quantifiable and interpretable way, not building the \textit{best} task method or network model over all of these possible modeling decisions. 

\subsection{Sensitivity to Network Density}
\label{subsec:comparison}
\begin{figure}
\centering
\includegraphics[width=\figwidthsmall\columnwidth]{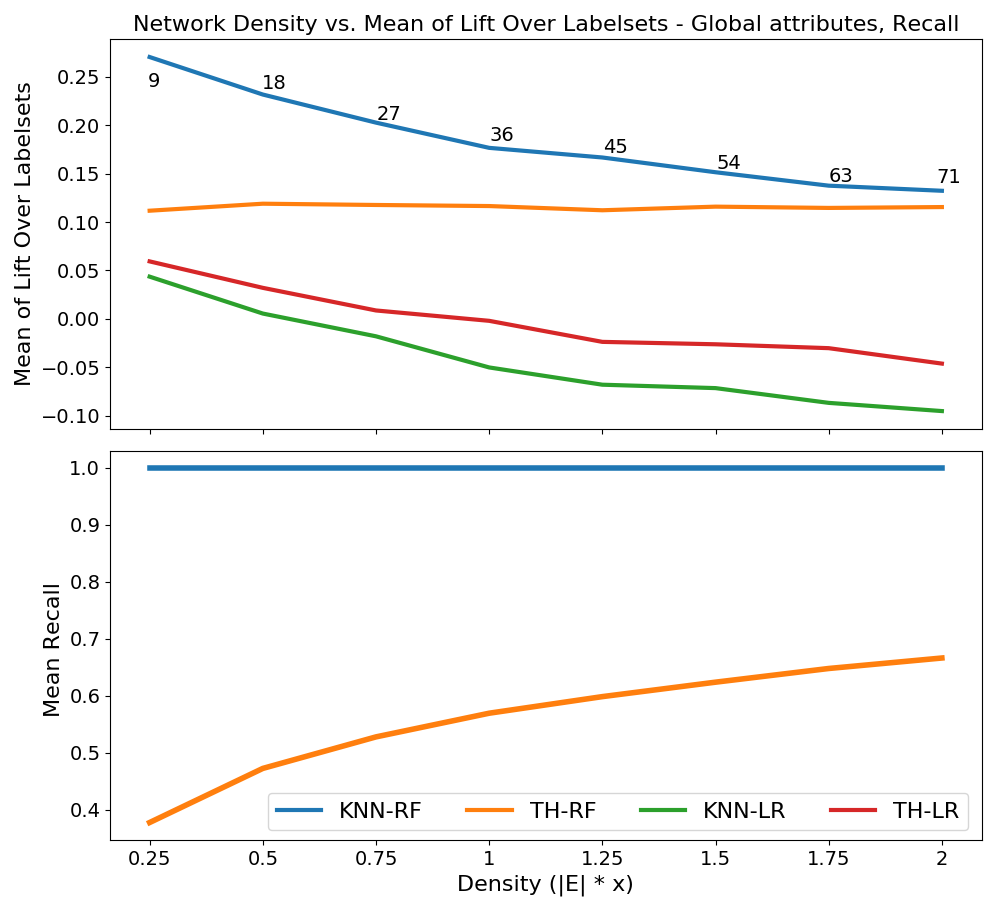}
\caption{(Top) Varying network model density as a factor of \lfm social network density ($x = 1$), vs. the mean of lift in precision over labelsets on the Global Attribute baseline, for model-classifier pairs. `k' for KNN given on the KNN-RF plot. (Bottom) Varying density vs. Recall: the fraction of nodes with positive labels with least one edge at `this' density (i.e. number of local method instances).} 
\label{fig:vary_density}
\end{figure}

Our evaluation of the \lfm social network fixes the density parameter $\lambda$ to infer other network models of equal density. We now evaluate how sensitive our inferred network models are to this choice. Figure \ref{fig:vary_density} (top) shows lift in precision against the Global Attributes baseline ($y$-axis) while varying density of network models ($x$-axis) where $x=1$ are at the density of $E_{20K}$. We show KNN and TH network models, and RF and LR classification methods (the first and second ranked against the GA baseline in Table \ref{tab:results_full}). On the KNN-RF plot, we label the actual `$k$' value at this $\lambda$. 

Figure \ref{fig:vary_density} (bottom) reports the mean over labelsets for the fraction of positive-label nodes in $L$ which learn a local classification method (i.e. over labelsets, the mean of positive-labeled nodes with at least one edge at `this' density). For example, on TH at $\lambda = 2|E_{200K}|$, each labelset on average builds classifiers for $\approx 0.7$ of positive-labeled nodes. Both classification methods on KNN and TH share a bottom plot. KNN ensures that all connected nodes in $G_{KNN}$ are allocated edges at any density, so nodes are not disconnected for any density value. 

Figure \ref{fig:vary_density} shows that more sparse network models perform better in all cases except TH-RF, which maintains performance over varying density. For the TH network model, the change in mean recall is small over the $x$-axis ($\approx0.25$). This suggests that with respect to TH-RF, additional edges tend to be associated with a small portion of the network and are uninformative to further classification. LR is a weaker method overall in our study, so TH-LR degrades slightly (but less than KNN-LR). On the KNN network, further edges are evenly allocated to nodes, breaking some local models where $k$ is too large. KNN converges to TH where the same high-degree preferential nodes in TH are robust to larger $k$ in KNN. Learning more sophisticated edge allocations from the attribute similarity distribution, such as varying $k$ subject to classifier performance could be more insightful to local properties of the network.

\subsection{Other Network Neighborhoods}

We measure the network models subject to a neighborhood function $\mathcal{N}(E, i)$. Above, we use simple network adjacency. However, further network traversal or some other function on the edge-set may be appropriate to evaluate `local' properties of interest. We formulate a neighborhood of variable size `k' as a \textit{node ordering} using breadth-first search (BFS) on the network topology, breaking ties at the same depth arbitrarily. Ordering nodes by BFS is an intuitive analogue to KNN because it respects non-decreasing order of geodesic distance. Similarly, varying `k' in KNN builds adjacency neighborhoods which respect the attribute similarity ordering under $\mathcal{S}()$.

\begin{figure}
\centering
\includegraphics[width=\figwidthsmall\columnwidth]{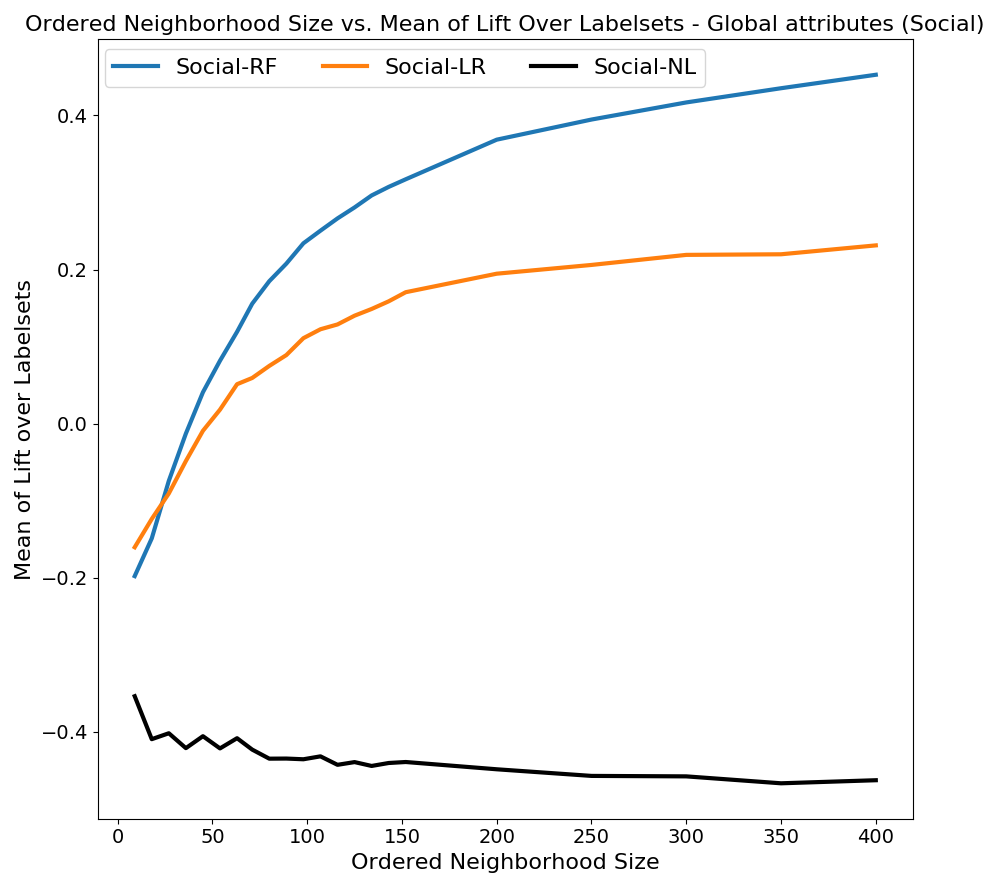}
\caption{Network neighborhood size using breadth-first search of variable size (x-axis) vs. the mean lift in precision over labelsets against the Global Attributes baseline (y-axis)} 
\label{fig:vary_neighborhood_depth}
\end{figure}

Figure \ref{fig:vary_neighborhood_depth} reports the mean lift in precision against the Global Attributes (GA) baseline ($y$-axis). We pass the method the `neighborhood' attributes and/or labels as previously, $\mathcal{C}(A_{\mathcal{N}(E, i)}, L_{\mathcal{N}(E, i)}, \vec{a}_i)$, where we form a neighborhood set of size `k' by order of node encounter using BFS traversal from $v_i$ ($x$-axis). 

This shows that performance continues to grow even to neighborhoods of size $400$, necessarily converging to the global model ($y=0$) at some larger $x$. Social-RF achieves parity with GA at $x=45$, and matches the best performance of KNN-RF (Figure \ref{fig:vary_density}) at $x=107$, while KNN achieves this performance with $k$=9. This shows that querying `similar' users even under our naive similarity is an order of magnitude more efficient in training cost. %although KNN is more sensitive to higher density.

This demonstrates that although we cannot use adjacency on the `friendship' network for learning local methods which perform better than global aggregates, this definition of `local' evaluates the quality of the network `near' the user and shows much better performance albeit at higher cost. Our node-ordering measures the efficiency of these local models in terms of the number of nodes in classification model training vs. the lift in precision.  

\subsection{Evaluating Labelsets}

Figure \ref{fig:heatmap} demonstrates considerable variability between labelset choice and local model performance. We want to quickly characterize or estimate the performance of a labelset by a relatively inexpensive computation such as our measure of network-label bias: i.e. $\mathcal{O}(n)$ in the number of edges.

\begin{figure}
\centering
\includegraphics[width=\figwidth\columnwidth]{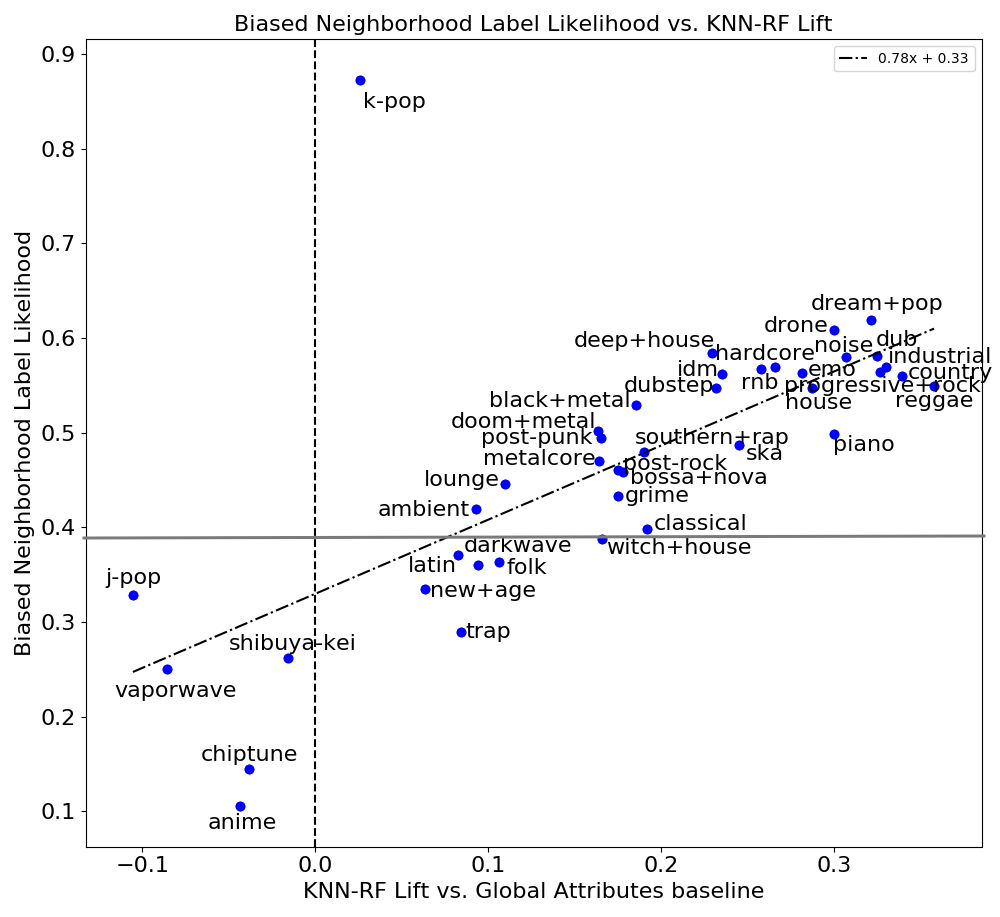}
\caption{The lift of precision for the KNN-RF model-classifier pair for each labelset against the Global Attribute (GA) baseline ($x$-axis), vs. the network-label bias $\mathcal{B}$ ($y$-axis). Points to the right of the dashed vertical line perform better than GA. The diagonal dashed line represents the linear model fit, showing a strong positive linear relationship between network-label bias and the final lift in precision. For a fixed network-label bias (e.g. $y = 0.4$, the horizontal line), points with a greater x value than the model perform better than expected by the model. perform better than expected, e.g. `witch+house', `piano', `reggae' all perform well.}
\label{fig:bias_v_perf}
\end{figure}

Figure \ref{fig:bias_v_perf} shows the lift in precision for KNN-RF on the Global Attributes (GA) baseline, over all labelsets ($x$-axis). This is against the network-label bias of each labelset ($y$-axis). A point to the right of the vertical dashed line performs better than the GA baseline. We fit a linear model, which shows a strong linear relationship. For a fixed $y$ value (e.g. the grey horizontal line), a point intersecting this line to the right of the model performs better than expected, e.g. `piano', `classical', and `witch+house' all have large $\Delta x$ from the model. `k-pop' greatly under-performs according to this linear model. 

\begin{figure}
\centering
\includegraphics[width=\figwidth\columnwidth]{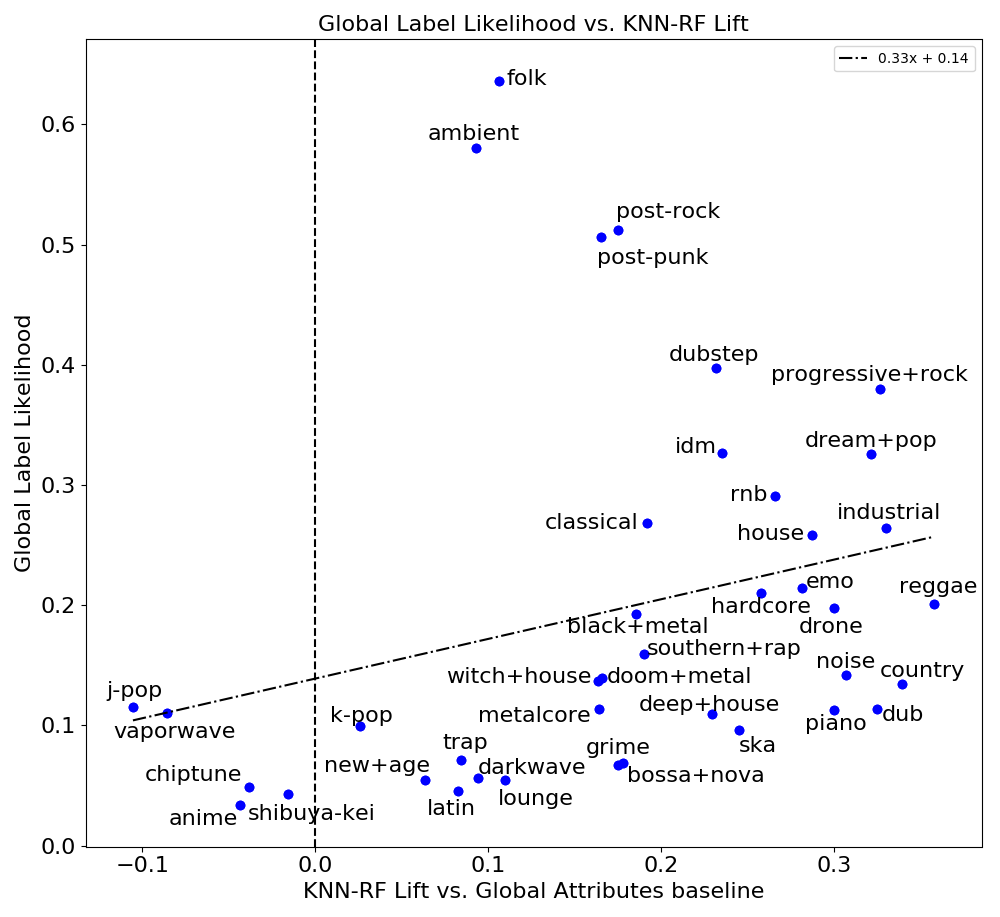}
\caption{The lift of precision for the KNN-RF model-classifier pair for each labelset against the Global Attribute (GA) baseline ($x$-axis), vs. the positive label prevalence: $\Pr(l_i = 1, l_i \in L)$, ($y$-axis). This plot shows a decreasing trend in performance after some prevalence. Well-performing labelsets include `piano,' `dub,' `country,' `noise' etc.}
\label{fig:likelihood_v_perf}
\end{figure}

Figure \ref{fig:likelihood_v_perf} shows a similar relationship between label likelihood and lift. We again report the lift in precision of KNN-RF against the Global Attributes (GA) baseline, over all labelsets ($x$-axis). This is against the positive-label prevalence: $|L|$. This suggests that labels above some prevalence (e.g. $ y > 0.25$) see diminishing returns in lift vs. the linear model. This is a very practical result since the number of local classification instances (i.e. runtime) grows with the positive-label prevalence. For example, `dub' and `country' labelsets provide 3 times the local learning compared to `folk' or `ambient,' but require 3-4 times fewer classification instances. 

Therefore, rare (but not too rare) and specialized (i.e. network biased) labelsets are the most effective for evaluating local learning in network models. This result matches our intuition; we are able to quantify it here.

%edge-count $\lambda = |E|$ (unless specified). For each label-set $L(`genre') \in L^*$, for each node $v_i \in V$, we learn a local classifier model on A and L (See Section \ref{subsec:classmodel}) subject to the neighborhood function $\mathcal{N}(i)$.  on each edge-set, and re-infer known label $l_i$. We report \emph{precision} over all nodes, yielding one value per combination of edge-set, label-set and classification model. (edge-set, label-set, classification model) tuple. We finally measure the \emph{lift} in precision, or the signed difference in precision for two models, e.g. $p_{rf} - p_{gf}$ (higher is better, with respect to the Random Forest model). Aggregating lift across any of these dimensions robustly summarizes the best-performing edge-set, or label-set, or classification model across many other problem configurations. We're particularly interested in the edge-set evaluation... \todo{continue here}

%We evaluate each network model on performing listener identification as an instance of collective classification from attributes and genre `listener' label-sets

% \todo{Figure running right now}
% \begin{figure}[h!]
% \centering
% \includegraphics[width=.99\columnwidth]{figs/density_th0}
% \caption{The lift for the two inferred graph models with Naive Bayes and Random Forest classification, where the network edge-count is a multiplicative factor of the \lfm social network: $\lambda = |E|x$, for label-sets with positive mean lift  }
% \label{fig:heatmap}
% \end{figure}
\section{Conclusion}
This work presented a general, modular framework for evaluating a given network as a representation for a set of tasks of interest. We instantiated a problem for collective node classification using simple, interpretable network and classification methods. We introduced simple and effective empirical measures of (1) network-label bias to estimate labelset performance, (2) global-attribute model baselines to quantify local network effects, and (3) variable-sized neighborhoods to compare performance across network models. 

There are several limitations of our current framework. We measured the observed network as-is, and compared against network models which may not correspond to actual social interactions. Constraining network models to local properties of the observed network may be more informative. For example, we may ensure that for a given node, KNN creates edges constrained by some geodesic distance (e.g. 1, yielding subgraphs of E), or some BFS search neighborhood. 

%We infer network models which are independent As we showed  First, our framework can only evaluate networks on enumerated network and task models, which can only quantify the local network properties over those selections. Furthermore, we would like to use other criteria to avoid enumeration of model pairs and the parameter-space.

Second, although we evaluate network inference robust over many labelsets, we don't demonstrate either robustness or specialization to different tasks. Future work will introduce a general, task-focused network inference model selection over further fundamental network tasks (e.g. link prediction). 

Third, our framework doesn't account for modeling cost in terms of classifier model size, number of local classifier instances, or actual encoding size in bytes. A further approach would constrain these costs, forcing models to reuse or select a small number of task method instances, evaluating on both task performance and parsimonious task method cost. 

%inferred networks which are robust to many labelsets network inference with respect to a single task, which may not be robust

%Second, our framework doesn't account for modeling cost in terms of task model size or number of local classifier instances. An alternative approach would constrain these task modeling costs to evaluate how well each network model \textit{summarizes} the attribute-label distributions subject to its topology. 

%minimize $k$ the number or size of local task models ($k << |V|$) on each network model, and use some model selection  

\section{Acknowledgements}

The work was in part supported by the National Science Foundation grants: III-1514126 (Berger-Wolf), CNS-1248080 (Berger-Wolf), CNS-1405886 (Kanich) and IGERT CNS-1069311 (Brugere).

\bibliographystyle{ACM-Reference-Format}
\bibliography{\jobname.bib,acmsmall-sample-bibfile}

% that's all folks
\end{document}